\documentclass{article} 
\usepackage{iclr2027_conference,times}
\usepackage{amsmath,amssymb}
\usepackage{amsthm}
\usepackage{booktabs}
\usepackage{multirow}
\usepackage{graphicx}
\usepackage{placeins}
\usepackage{algorithm}
\usepackage{algpseudocode}

\usepackage{amsmath,amsfonts,bm}

\def\eqref#1{equation~\ref{#1}}

\def\1{\bm{1}}

\DeclareMathAlphabet{\mathsfit}{\encodingdefault}{\sfdefault}{m}{sl}
\SetMathAlphabet{\mathsfit}{bold}{\encodingdefault}{\sfdefault}{bx}{n}

\usepackage[hidelinks]{hyperref}
\usepackage{url}

\providecommand{\Rj}{R_j}
\providecommand{\Hj}{H_j}

\providecommand{\OPT}{\mathrm{OPT}}
\providecommand{\argmaxop}{\operatorname*{arg\,max}}
\providecommand{\ind}[1]{\mathbf{1}\!\left[#1\right]}
\providecommand{\fcov}{f_{\mathrm{cov}}}
\providecommand{\method}{\textsc{QESChunker}}

\newtheorem{proposition}{Proposition}
\newtheorem{theorem}{Theorem}

\title{QESChunker: A Single Objective Unifies\\ Overlapping and Non-Overlapping Chunking for RAG}

\author{\parbox{\dimexpr\textwidth-2\tabcolsep\relax}{\raggedright\normalfont\small
\textit{Yifan Zhao\textsuperscript{1}, Qianyang Zhao\textsuperscript{1},
Xinyu Guan\textsuperscript{1}, kai wei\textsuperscript{2}, Yuming Deng\textsuperscript{1}}\\[4pt]
\textsuperscript{1}Alibaba Group (alibaba-inc.com)\\
\textsuperscript{2}Taobao (China) Software Co., Ltd. (taobao.com)
}}

\iclrfinalcopy 

\begin{document}
\raggedbottom

\maketitle
\lhead{} 
\renewcommand{\headrulewidth}{0pt}

\begin{abstract}
Chunking is a fundamental preprocessing step in retrieval-augmented generation (RAG), yet most existing chunkers rely primarily on heuristics. We introduce the Question--Evidence-Supervised Chunker (QESChunker), to our knowledge the first method for constructing static chunk indices from question--evidence supervision. QESChunker constructs document-specific chunk indices by optimizing over human-annotated or LLM-generated question--evidence pairs, without access to held-out test queries. We formulate non-overlapping and overlapping index construction through a unified question--evidence utility-maximization objective, turning chunking into a supervised combinatorial optimization problem. Under the non-overlap constraint, the objective decomposes over chunks and can be optimized exactly in polynomial time by dynamic programming. When overlap is allowed, the objective becomes monotone submodular; exact optimization is NP-hard, but our greedy achieves a $(1-1/\sqrt{e})$-approximation. Across two self-constructed and three public datasets, both variants outperform existing chunking methods in complete-evidence recall in nearly all evaluated settings, under the same overlap regime and a fixed hybrid retriever. These results recast chunking as supervised, budget-aware index construction for evidence retrieval.
\end{abstract}

\section{Introduction}
\label{sec:intro}

Retrieval-augmented generation (RAG) conditions a language model on passages retrieved from an external corpus~\citep{lewis2020rag,gao2024ragsurvey}. Before any retrieval can occur, each document must be segmented into \emph{chunks} that are embedded and indexed. Where the chunk boundaries fall largely determines whether the evidence for a question can be retrieved \emph{completely}.

Despite its importance, chunking is overwhelmingly heuristic in practice. The baselines evaluated in this paper span fixed-length and fixed-overlap passages used by retrieval systems~\citep{lewis2020rag,guu2020realm,karpukhin2020dpr}, natural paragraphs, lexical-cohesion segmentation with TextTiling~\citep{hearst1997texttiling}, embedding-based ClusterSemanticChunker~\citep{smith2024chunking}, LLM-predicted boundaries with LumberChunker~\citep{stlaurent2024lumberchunker}, sentence-window retrieval~\citep{llamaindex_sentence_window}, and hierarchical Parent-Document and Auto-Merging retrieval~\citep{llamaindex_auto_merging}. These practical alternatives cover non-overlapping, overlapping, semantic, LLM-based, and hierarchical designs, but none optimizes an explicit question--evidence objective.

\begin{figure}[t]
\centering
\includegraphics[width=0.82\linewidth,trim=55 100 20 60,clip]{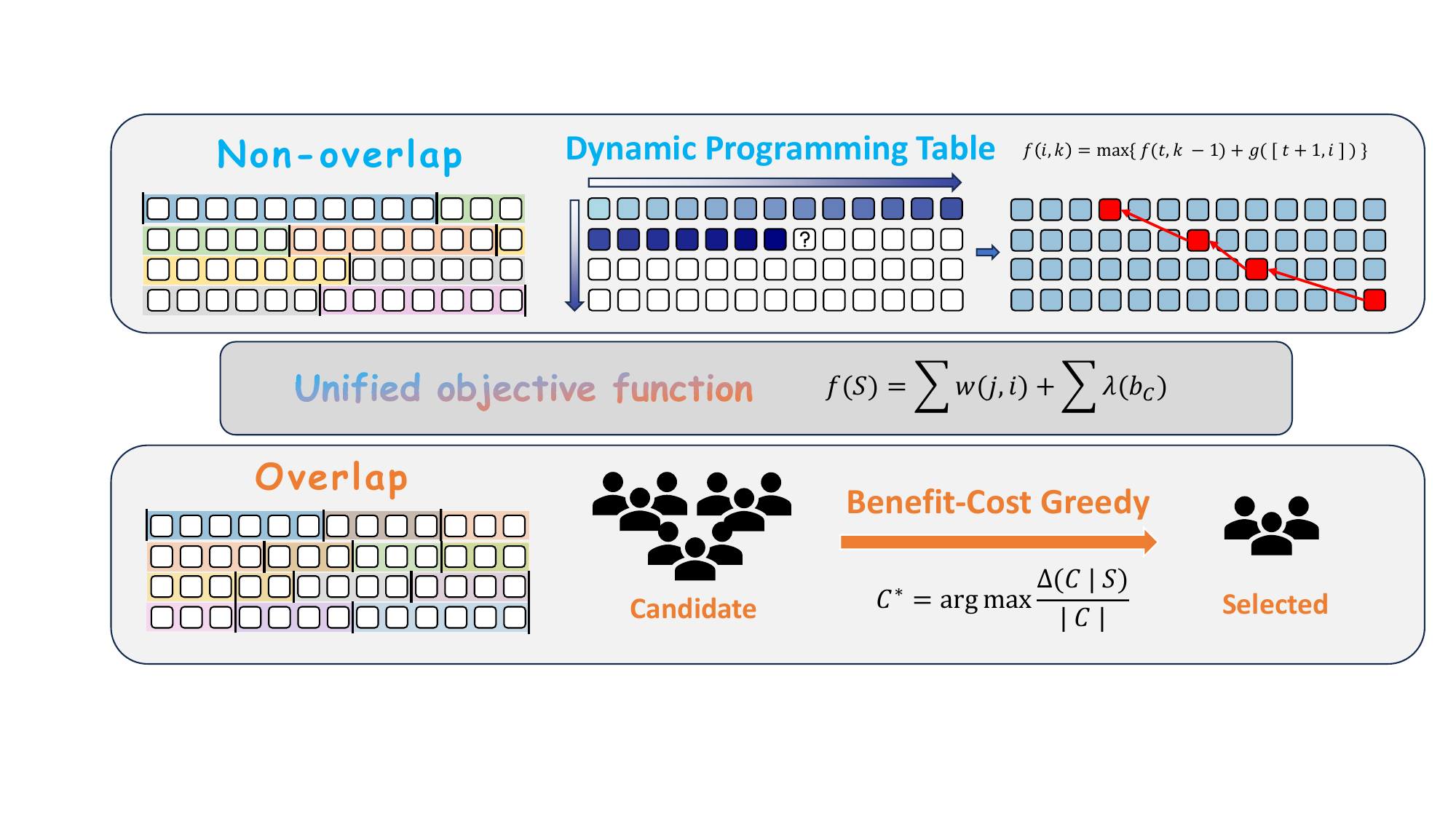}
\caption{Overview of \method{}. A shared question--evidence objective supports exact dynamic programming for non-overlapping partitions and benefit--cost greedy selection for overlapping indices.}
\label{fig:architecture}
\end{figure}

We cast chunking as an explicit optimization problem and show that a \emph{single} objective, subjected to two different structural constraints, yields two algorithms with complementary guarantees. We call the resulting framework \method{} (\textbf{Q}uestion--\textbf{E}vidence \textbf{S}upervised Chunker): questions and their evidence blocks supervise where boundaries should fall, and the same objective is then solved exactly or near-optimally under each constraint. Figure~\ref{fig:architecture} summarizes the framework.

Our contributions are:

\begin{enumerate}
    \item \textbf{A unified objective.} We model a document as a sequence of atomic blocks and define a two-term objective comprising a coverage term and a structural term (\S\ref{sec:objective}).

    \item \textbf{Two algorithms under different constraints.} Under a \emph{non-overlap} constraint---that is, a contiguous partition into multiple chunks---the objective becomes additive over chunks, and an exact dynamic program returns the global optimum in polynomial time (\S\ref{sec:dp}). Under an \emph{overlap} constraint, an index budget over a candidate chunk family must be satisfied. We prove that exact optimization of the objective is \textbf{NP-hard} by a reduction from Grid Segment Covering (\S\ref{sec:greedy}), and give a benefit-cost greedy algorithm that attains a $(1-1/\sqrt{e})\approx0.393$ guarantee for budgeted monotone submodular maximization under the corrected analysis of \citet{tang2021modified}.

    \item \textbf{Broad wins on multiple datasets.} Across two self-constructed Chinese policy corpora and three public English corpora (QASPER, PeerQA, and PrivacyQA), under a fixed hybrid retriever, our non-overlap DP ranks first in $9/10$ corpus$\times$tier display cells and our overlap method in $10/10$ (Tables~\ref{tab:main_nonoverlap} and~\ref{tab:main_overlap}).
\end{enumerate}

\section{Related Work}
\label{sec:related}

\textbf{Chunking and adjacent segmentation.} Our evaluated baselines are fixed-length, natural-paragraph, TextTiling, ClusterSemanticChunker, LumberChunker, fixed-window, Sentence-Window, Parent-Document, and Auto-Merging. Classical topic and discourse segmenters such as C99 and supervised neural boundary models optimize segmentation labels rather than RAG evidence recall~\citep{choi2000c99,koshorek2018,lukasik2020}. Other adjacent systems change more than chunk boundaries: Dense X replaces passages with generated propositions, Late Chunking changes contextual embedding computation, and RAPTOR introduces recursive summaries and hierarchical retrieval~\citep{chen2024densex,gunther2024latechunking,sarthi2024raptor}. We therefore exclude them from matched comparisons that isolate boundary selection under a fixed retriever and original-atom budget.

\textbf{Submodular optimization.} Our overlap solver draws on classical monotone-submodular maximization~\citep{nemhauser1978,krause2014}; the corresponding non-overlapping partition instead admits an exact dynamic program.

\section{Problem Setup and Unified Objective}
\label{sec:objective}

\subsection{Atomic blocks, evidence, and hits}

A document is atomized into a sequence $u_1,\dots,u_n$ of \emph{atomic blocks} by splitting on minimal punctuation; blocks are the common accounting unit for both indexing cost and returned context. Let $B\subseteq[n]$ be the set of block indices that terminate natural paragraphs. We are given $m$ fitting questions. For question $j$:
\begin{itemize}
    \item $\Rj$ is the set of gold evidence atomic blocks;
    \item $\Hj$ is the \emph{hit set}: the top-$\kappa$ atomic blocks most similar to the question embedding, i.e., the blocks that a first-stage retriever would actually surface.
\end{itemize}
Let $S$ denote a set of chunks (either a partition or an overlapping set). A chunk $C=[a,b]$ covers blocks $a,\dots,b$.

\subsection{Hit-gated weighted coverage with a tiered structural prior}

Treat each pair ``question $j$, evidence block $r$'' as an element to be covered, with weight
\begin{equation}
w(j,r)=\frac{1}{m\,\lvert \Rj\rvert},\qquad \sum_{j}\sum_{r\in \Rj} w(j,r)=1.
\label{eq:weights}
\end{equation}
A single chunk $C$ covers the element set
\begin{equation}
\Gamma(C)=\big\{(j,r):\; r\in C\cap \Rj,\ \ C\cap \Hj\neq\varnothing\big\}.
\label{eq:gamma}
\end{equation}
The \emph{hit gate} $C\cap\Hj\neq\varnothing$ encodes retrieval reality: a chunk that contains none of question $j$'s hit blocks will not be surfaced for $j$, so it cannot contribute coverage for $j$ no matter how much gold text it holds.

The unified objective adds a tiered structural prior over chunk right-endpoints $b_C$:
\begin{equation}
\boxed{\;f(S)=\underbrace{\sum_{(j,r)\,\in\,\bigcup_{C\in S}\Gamma(C)} w(j,r)}_{\text{coverage term } }\;+\;\underbrace{\sum_{C\in S}\lambda(b_C)}_{\text{structural term}}\;}
\label{eq:objective}
\end{equation}
where a chunk's structural term is determined by the terminal punctuation of its right block:
\begin{equation}
\lambda(b_C)=\lambda \times
\begin{cases}
\rho_1, & b_C\text{ ends a paragraph},\\
\rho_2, & b_C\text{ ends a sentence but not a paragraph},\\
\rho_3, & b_C\text{ ends a clause (semicolon or colon)},\\
\rho_4, & b_C\text{ ends at a comma or another position},
\end{cases}
\qquad
\rho_1\ge\rho_2\ge\rho_3\ge\rho_4\ge0.
\label{eq:tiers}
\end{equation}
The parameter $\lambda$ controls the contribution of the structural term. In this paper, the tier weights $\rho_1,\rho_2,\rho_3,\rho_4$ are set to $0.1$, $0.005$, $0.001$, and $0$, respectively.

\textbf{Why hierarchical boundary rewards.} Paragraph endings usually mark the strongest semantic boundaries; when a cut must fall inside a paragraph, sentence endings are preferable to clause endings, and clause endings are preferable to arbitrary positions. The tier weights encode this hierarchy while allowing question--evidence supervision to trade structural preference against evidence coverage.

\subsection{Partition decomposition}

When $S$ is a \emph{partition}, its chunks are disjoint, so $\bigcup_{C\in S}\Gamma(C)$ is a disjoint union and $f$ decomposes as a sum of per-chunk values $g(C)$---exactly the condition that makes dynamic programming valid (\S\ref{sec:dp}). When overlap is allowed this decomposition breaks and $f$ must be treated as a set function, which is where submodularity earns its keep (\S\ref{sec:greedy}). Both settings therefore share the same $f$ and, under the evaluation convention of \S\ref{sec:setup}, directly comparable results.

\section{Setting 1: Non-overlap Partition}
\label{sec:dp}

Given the prior parameters $K$, $L_{\min}$, and $L_{\max}$, we seek a contiguous partition of $u_1,\ldots,u_n$ into exactly $K$ chunks such that every chunk $C$ satisfies $L_{\min}\le\lvert C\rvert\le L_{\max}$ or is admitted by the whole-paragraph exception below:
\begin{equation}
\max_{S}\ f(S)
\quad\text{s.t.}\quad
S\text{ is a contiguous }K\text{-partition satisfying the length rule.}
\label{eq:partition}
\end{equation}

\textbf{Whole-paragraph exception.} Because a natural paragraph is often a semantic unit, an entire paragraph longer than $L_{\max}$ may form one chunk; oversized chunks that cross paragraph boundaries remain forbidden.

For every feasible chunk $C$, define its per-chunk utility as
\begin{equation}
g(C)=\sum_{(j,r)\in\Gamma(C)}w(j,r)+\lambda(b_C).
\label{eq:chunk_gain}
\end{equation}
Let $f(i,k)$ be the maximum utility of a feasible partition of the prefix $u_1,\ldots,u_i$ into exactly $k$ chunks, with value $-\infty$ when no such partition exists. The sequential structure gives the recurrence
\begin{equation}
f(i,k)=\max_{t\,\in\,\mathcal{T}(i)}\Big\{f(t,k-1)+g\big([t{+}1,i]\big)\Big\},\qquad f(0,0)=0,
\label{eq:dp}
\end{equation}
where the predecessor set adjoins the whole-paragraph exception to the usual length window:
\begin{equation}
\begin{aligned}
\mathcal{T}(i)={}&\underbrace{\{t:\ i-L_{\max}\le t\le i-L_{\min},\ t\ge 0\}}_{\text{length window}}\\
&{}\cup\underbrace{\big\{\mathrm{start}(i)-1\ \big|\ i\in B,\ i-\mathrm{start}(i)+1>L_{\max}\big\}}_{\text{whole-paragraph exception}}.
\end{aligned}
\label{eq:predecessors}
\end{equation}
with $\mathrm{start}(i)$ the first block of the paragraph ending at $i$. The optimum is $f(n,K)$; cut points are recovered by back-pointers.

\textbf{Time complexity analysis.} Let $W=L_{\max}-L_{\min}+1$. After interval gains are computed incrementally with block-to-question inverted lists, each DP state examines at most $W$ length-feasible predecessors plus one whole-paragraph predecessor. The running time is therefore $O(nKW)$ and the table with back-pointers uses $O(nK)$ space.

\begin{algorithm}[H]
\caption{Exact DP for non-overlapping chunking}
\label{alg:dp}
\begin{algorithmic}[1]
\State \textbf{Input:} $u_{1:n}$, $(\Rj,\Hj)_{j=1}^m$, $K$, $L_{\min}$, $L_{\max}$, $B$
\State Initialize $f(0,0)\gets0$ and all other states to $-\infty$.
\For{$i=1,\dots,n$}
    \State Form $\mathcal T(i)$ and compute $g([t{+}1,i])$ incrementally for $t\in\mathcal T(i)$.
    \For{$k=1,\dots,K$}
        \State Update $f(i,k)$ by \eqref{eq:dp} and store its maximizing predecessor.
    \EndFor
\EndFor
\State \Return the partition obtained by backtracking from $f(n,K)$.
\end{algorithmic}
\end{algorithm}

\section{Setting 2: Overlap Selection}
\label{sec:greedy}

Given a prescribed index budget $B_{\text{index}}$, which caps the total number of indexed atomic blocks, and prior length bounds $L_{\min}$ and $L_{\max}$, let
\[
\mathcal V=\{[a,b]:1\le a\le b\le n,\ L_{\min}\le b-a+1\le L_{\max}\}
\]
be the candidate family. Allowing overlap, the index is a set $S\subseteq\mathcal V$ rather than a partition, and we solve
\begin{equation}
\max_{S\subseteq \mathcal V}\ f(S)\quad\text{s.t.}\quad \sum_{C\in S}\lvert C\rvert\le B_{\text{index}}.
\label{eq:knapsack}
\end{equation}
Here each selected chunk incurs cost $\lvert C\rvert$, and the structural prior is instantiated as a paragraph-end reward (the sentence and clause tiers are zero). Unlike the partition setting, the overlapping candidate family uses only the prescribed length bounds and does not invoke the whole-paragraph exception.

\subsection{Monotone submodularity}

\begin{proposition}[Monotone submodularity]
\label{prop:submodular}
The set function $f$ in \eqref{eq:objective} is monotone and submodular over chunks drawn from any candidate family.
\end{proposition}
\begin{proof}[Proof sketch]
The coverage term is a weighted coverage function: each element $(j,r)$ contributes $w(j,r)$ once, the first time it is covered by a selected chunk. Weighted coverage is monotone submodular~\citep{nemhauser1978}. The prior term is non-negative and modular because each summand depends only on one chunk's right endpoint. Their sum is therefore monotone submodular. Appendix~\ref{app:proof} gives the full proof.
\end{proof}


%

\subsection{Exact optimization is NP-hard}

We set all structural-prior weights to zero, so the result applies directly to
the coverage term of \eqref{eq:objective} and implies hardness of the full
objective when the prior weights are allowed to be zero.

\begin{theorem}[NP-hardness for contiguous chunks]
\label{thm:nphard}
Let the maximum chunk length $L$ be part of the input and let
\[
  \mathcal V_L
  =\{[a,b]:1\le a\le b\le n,\ b-a+1\le L\}.
\]
Deciding whether there exists $S\subseteq\mathcal V_L$ with
$\sum_{C\in S}|C|\le B_{\mathrm{index}}$ and $\fcov(S)=1$ is NP-complete,
even when $|R_j|=1$ and $|H_j|=2$ for every question, the weights are
$w(j,r)=1/m$, and all structural-prior weights are zero.
Consequently, exact maximization of the coverage-only objective over
contiguous chunks is NP-hard. The same conclusion holds when
$\mathcal V_L$ consists of all intervals of length \emph{exactly} $L$.
\end{theorem}

\begin{proof}[Proof sketch]
We reduce from \textsc{Grid Segment Covering}, which is NP-complete by
\citet[Theorem~7]{bergren2020segments}. Its input is a set of $m$ vertical
unit segments $e_j=[(x_j,y_j),(x_j,y_j+1)]$ on an integer grid of side
$q=O(m)$ and an integer $k$; a horizontal length-$2$ segment covers $e_j$
if it contains either endpoint. We may assume $1\le k\le m$.

Set $D=q+2m+3$ and encode $(x,y)$ by $Dy+x+3$. The endpoints of $e_j$ then
become $\{p_j,p_j+D\}$, where $p_j=Dy_j+x_j+3$. Encoded points on distinct
rows are more than $2$ apart, so the grid instance is coverable by $k$
horizontal length-$2$ segments exactly when $k$ length-$2$ intervals on the
line contain at least one point from every pair $\{p_j,p_j+D\}$. Such line
intervals may be assumed to have integer left endpoints.

Let $p_{\max}=\max_j p_j$ and construct $n=p_{\max}+2D+2$ atomic blocks.
For each $j$, define
\begin{equation}
  R_j=\{p_j+D\},\qquad H_j=\{p_j,p_j+2D\}.
  \label{eq:reduction}
\end{equation}
Give its sole evidence pair weight $1/m$, set all prior weights to zero,
and choose $L=D+3$ and $B_{\mathrm{index}}=kL$. For a length-$L$ chunk
$C_t=[t,t+D+2]$,
\begin{align*}
  (j,p_j+D)\in\Gamma(C_t)
  &\iff t\in[p_j-2,p_j]\ \cup\ [p_j+D-2,p_j+D] \\
  &\iff [t,t+2]\cap\{p_j,p_j+D\}\ne\varnothing.
\end{align*}
Thus $C_t$ covers question $j$ exactly when the corresponding length-$2$
interval covers an encoded endpoint of $e_j$.

This correspondence maps every grid cover of size at most $k$ to a feasible
chunk family with $\fcov=1$. Conversely, every chunk with nonempty coverage
spans an evidence block and a hit block $D$ positions apart, and hence costs
at least $D+1$. Since $(k+1)(D+1)-kL=D+1-2k>0$, a feasible solution contains
at most $k$ such chunks. Extending each to length $L$ preserves coverage,
after which the same correspondence yields a grid cover of size at most
$k$. Appendix~\ref{app:nphard} gives both directions in full, including the
normalization and boundary arguments.

The construction is polynomial because $D,L=O(m)$, $n=O(m^2)$, and
$|\mathcal V_L|=O(nL)=O(m^3)$; feasibility and coverage are also
polynomially verifiable. This proves NP-completeness. If only length-$L$
chunks are allowed, the same construction applies and the budget directly
limits the solution to $k$ chunks, proving the final claim.
\end{proof}

\subsection{A \texorpdfstring{$(1-1/\sqrt{e})$}{(1-1/sqrt(e))} cost-benefit greedy}

By Proposition~\ref{prop:submodular}, $f$ is monotone submodular, so \eqref{eq:knapsack} is budgeted monotone submodular maximization, for which the \emph{cost-benefit greedy with a singleton safeguard} (\textsc{MGreedy}) attains the following conservative guarantee under the corrected analysis of \citet{tang2021modified}.

\begin{theorem}[Approximation guarantee~\citep{tang2021modified}]
\label{thm:greedy}
Let $f$ be monotone submodular over $\mathcal V$ with $f(\varnothing)\ge0$ and $\OPT=\max\{f(S):S\subseteq\mathcal V,\ \sum_{C\in S}\lvert C\rvert\le B_{\text{index}}\}$. Algorithm~\ref{alg:greedy} returns $S$ with
\begin{equation}
f(S)\ \ge\ \Bigl(1-\tfrac{1}{\sqrt{e}}\Bigr)\,\OPT\ \approx\ 0.393\,\OPT .
\end{equation}
\end{theorem}

\citet{tang2021modified} close a gap in the earlier analysis of \citet{khuller1999} and prove the stronger $0.405$ approximation factor for the same modified greedy algorithm; we retain the weaker $(1-1/\sqrt{e})$ bound here. Each round selects the interval with maximum marginal gain per unit cost,
\begin{equation}
C^\star=\argmaxop_{C\in\mathcal V}\ \frac{\Delta(C\mid S)}{\lvert C\rvert},\qquad
\Delta(C\mid S)=f(S\cup\{C\})-f(S),
\label{eq:density}
\end{equation}
and the \emph{singleton safeguard} compares the greedy's result with the best single feasible chunk, returning the better. Reaching the $(1-1/e)\approx0.632$ guarantee instead requires enumerating all initial subsets of size $\le3$~\citep{sviridenko2004}, which is $\lvert\mathcal V\rvert^3\approx10^{14}$ at our scale and infeasible; and for coverage objectives under cardinality constraints no polynomial algorithm can beat $(1-1/e)$ unless $\mathrm{P}=\mathrm{NP}$~\citep{feige1998}, so the residual $\sqrt e$ gap is the price of tractability. \emph{Lazy evaluation} makes the greedy practical: by submodularity a cached marginal gain is an upper bound on the current one, so a max-heap of stale densities needs only tens of exact re-evaluations per accepted chunk, and it returns exactly the same set as the eager cost-benefit greedy. Algorithm~\ref{alg:greedy} summarizes the procedure.

\begin{algorithm}[t]
\caption{Lazy cost-benefit greedy with singleton safeguard}
\label{alg:greedy}
\begin{algorithmic}[1]
\State \textbf{Input:} $\mathcal V$, $B_{\text{index}}$, and objective $f$
\State $S\gets\varnothing$; heapify $\mathcal V$ by $\Delta(C\mid S)/\lvert C\rvert$.
\While{the heap is non-empty and $\sum_{C\in S}\lvert C\rvert<B_{\text{index}}$}
    \State Pop $C$ and recompute $\Delta(C\mid S)/\lvert C\rvert$.
    \If{$C$ is feasible, has positive gain, and remains heap-maximal}
        \State $S\gets S\cup\{C\}$.
    \ElsIf{$C$ is feasible and $\Delta(C\mid S)>0$}
        \State Reinsert $C$ with its updated density.
    \Else
        \State Discard $C$.
    \EndIf
\EndWhile
\State $s^\star\gets\argmaxop_{C\in\mathcal V:\lvert C\rvert\le B_{\text{index}}}f(\{C\})$.
\State \Return $\argmaxop_{T\in\{S,\{s^\star\}\}} f(T)$.
\end{algorithmic}
\end{algorithm}

In practice, the empirical gap between the algorithm's value and the optimum is generally much smaller than $1 - \dfrac{1}{\sqrt{e}}$.

\section{Experiments}
\label{sec:experiments}

\subsection{Experimental Setup}
\label{sec:setup}

\textbf{Datasets and supervision.} Table~\ref{tab:datasets} summarizes the corpora and supervision protocols.

\begin{table}[H]
\caption{Dataset statistics and supervision.}
\label{tab:datasets}
\centering
\small
\setlength{\tabcolsep}{4pt}
\begin{tabular*}{\textwidth}{@{\extracolsep{\fill}}llrrrr}
\toprule
\textbf{Dataset} & \textbf{Type} & \textbf{Docs} & \textbf{Blocks} & \textbf{Paragraphs} & \textbf{Questions} \\
\midrule
Party Congress report & Self-built & 1 & 2,842 & 147 & 642 \\
Public-health white paper & Self-built & 1 & 2,884 & 232 & 683 \\
QASPER~\citep{dasigi2021qasper} & Public & 185 & 103,024 & 12,548 & 532 \\
PeerQA~\citep{peerqa2025} & Public & 70 & 43,656 & 6,222 & 136 \\
PrivacyQA~\citep{ravichander2019privacyqa} & Public & 34 & 4,931 & 4,931 & 1,276 \\
\bottomrule
\end{tabular*}
\end{table}

\textbf{Supervision protocol.} Both self-built corpora use eight-fold cross-validation, with boundary fitting and testing separated in every fold. For each public document containing $n$ atomic blocks, the \texttt{qwen3.8-max} API generates $n/5$ synthetic questions exclusively for boundary fitting; contiguous-evidence and cross-paragraph-evidence questions are generated in a $1{:}1$ ratio. Evaluation then uses only the datasets' native public questions and evidence annotations. We compare against the non-overlap and overlap baseline families described in \S\ref{sec:related}, using identical retrieval and returned-context budgets.

Automatic checks validate the generated fitting data. PrivacyQA treats every atom as a paragraph, so the natural-paragraph baseline is not applicable.

\textbf{Evaluation convention.} $@k$ denotes complete-evidence recall when the deduplicated returned context contains $k$ atomic blocks. $@k\%$ denotes the same metric when the context budget is $k\%$ of the document's atomic blocks.

All methods use the same embedding--BM25 retriever with section-title prefixes. Coverage is aggregated per question; self-built results are averaged over the eight folds. The parameter study below fixes operating points before the cross-corpus comparison.

\FloatBarrier
\subsection{Parameter Sweeps}
\label{sec:sweeps}

We vary one factor at a time on the public-health white paper while holding the retriever, returned-atom budget, and eight folds fixed.

\subsubsection{Structural weight}

\begin{figure}[H]
\centering
\includegraphics[width=0.68\textwidth]{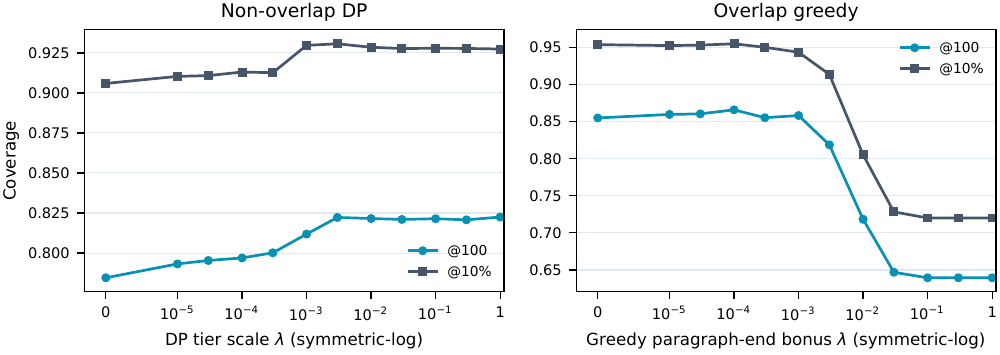}
\caption{Coverage over the $\lambda$ sweep at $@100$ and $@10\%$.}
\label{fig:covid_lambda_sweep}
\end{figure}

Figure~\ref{fig:covid_lambda_sweep} shows complementary trends. For DP, coverage rises as small positive structural weights are introduced and then reaches a broad plateau around $\lambda=10^{-2}$; both display budgets follow the same pattern. Greedy is stable near zero through small bonuses but declines once $\lambda$ becomes large, sharply so at $\lambda\ge10^{-2}$, indicating that an over-weighted paragraph prior displaces evidence-bearing chunks.

\subsubsection{Length geometry}

\begin{figure}[H]
\centering
\includegraphics[width=0.68\textwidth]{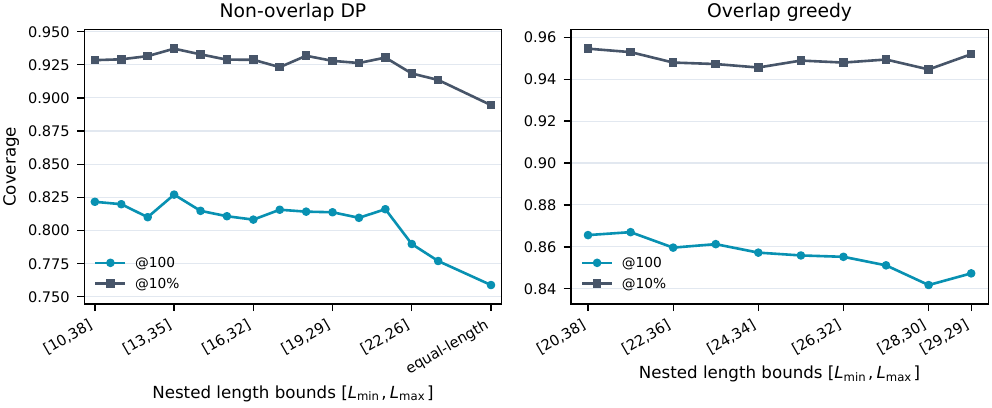}
\caption{Coverage as candidate-length intervals contract.}
\label{fig:covid_length_bounds_sweep}
\end{figure}

Figure~\ref{fig:covid_length_bounds_sweep} shows that both variants change smoothly as the feasible interval contracts rather than exhibiting a narrow optimum. DP is nearly flat over broad ranges and degrades only near the fixed-length limit; greedy is similarly stable at $@10\%$, with a modest decline at $@100$. At the rightmost point $L_{\min}=L_{\max}$, so the admissible chunks are exactly those of equal-length splitting and the two settings are fully equivalent.

\subsubsection{Whole-paragraph exception}

Table~\ref{tab:pararelax} shows that the whole-paragraph exception helps at small returned-context budgets but provides no advantage at large budgets. For fairness in the subsequent comparison, we enable the exception in every non-overlap main experiment.

\begin{table}[H]
\caption{White-paper coverage with the whole-paragraph exception disabled or enabled; bold marks the better result.}
\label{tab:pararelax}
\centering
\scriptsize
\setlength{\tabcolsep}{3pt}
\begin{tabular*}{\textwidth}{@{\extracolsep{\fill}}l*{11}{c}}
\toprule
& \multicolumn{5}{c}{\textbf{Tight budgets (atoms)}} & \multicolumn{6}{c}{\textbf{Larger budgets (atoms)}} \\
\cmidrule(lr){2-6}\cmidrule(lr){7-12}
\textbf{Exception} & 50 & 58 & 100 & 144 & 200 & 288 & 305 & 433 & 577 & 721 & 865 \\
\midrule
Disabled & 0.7076 & 0.7361 & 0.8142 & 0.8598 & 0.8972 & \textbf{0.9310} & \textbf{0.9378} & \textbf{0.9595} & \textbf{0.9714} & \textbf{0.9828} & \textbf{0.9863} \\
Enabled & \textbf{0.7177} & \textbf{0.7490} & \textbf{0.8216} & \textbf{0.8631} & \textbf{0.8980} & 0.9285 & 0.9340 & 0.9587 & 0.9703 & 0.9798 & 0.9853 \\
Enabled $-$ disabled (pp) & $+1.01$ & $+1.29$ & $+0.73$ & $+0.33$ & $+0.08$ & $-0.25$ & $-0.38$ & $-0.08$ & $-0.12$ & $-0.29$ & $-0.10$ \\
\bottomrule
\end{tabular*}
\end{table}

\FloatBarrier
\subsection{Main Experiments}
\label{sec:main}

Based on \S\ref{sec:sweeps}, we fix all operating points before the cross-corpus comparison. For non-overlap, we set $\kappa=5$, $\lambda=0.01$, take $[L_{\min},L_{\max}]$ as the 10th--90th percentiles of paragraph length, and set $K$ to the number of paragraphs. For overlap, we set $\lambda=10^{-4}$, use the same percentile-based length bounds, and set the total budget to three times the document length. We then compare non-overlapping methods under a common partition constraint and overlapping methods under the same returned-context convention.

\textbf{Residual-budget concentration (overlap only).} When the primary coverage objective saturates while index budget remains, we use
\begin{equation}
f_2(S)=\frac{1}{m}\sum_{j}\ \max_{C\in S,\ C\cap\Hj\neq\varnothing}\ \frac{\lvert C\cap \Rj\rvert}{\lvert \Rj\rvert},
\label{eq:concentration}
\end{equation}
and continue selection with $f+\delta f_2$, where $\delta=0.05$, which favors candidates that concentrate more of a question's evidence in one retrievable chunk.

\begin{table}[H]
\caption{\textbf{Non-overlap setting:} complete-evidence coverage at the two display tiers ($@100$ fixed atoms; $@10\%$ of document atoms). On PrivacyQA the natural-paragraph baseline is degenerate (one atomic block per paragraph) and is therefore marked ``--''.}
\label{tab:main_nonoverlap}
\centering
\scriptsize
\setlength{\tabcolsep}{3pt}
\begin{tabular*}{\textwidth}{@{\extracolsep{\fill}}l cc cc cc cc cc}
\toprule
& \multicolumn{2}{c}{\textbf{Congress report}} & \multicolumn{2}{c}{\textbf{White paper}} & \multicolumn{2}{c}{\textbf{QASPER}} & \multicolumn{2}{c}{\textbf{PeerQA}} & \multicolumn{2}{c}{\textbf{PrivacyQA}} \\
\cmidrule(lr){2-3}\cmidrule(lr){4-5}\cmidrule(lr){6-7}\cmidrule(lr){8-9}\cmidrule(lr){10-11}
\textbf{Method} & @100 & @10\% & @100 & @10\% & @100 & @10\% & @100 & @10\% & @100 & @10\% \\
\midrule
\textbf{\method{} (exact DP)} & \textbf{0.9025} & \textbf{0.9781} & \textbf{0.8216} & \textbf{0.9285} & \textbf{0.7588} & \textbf{0.6029} & 0.7341 & \textbf{0.6589} & \textbf{0.8687} & \textbf{0.3166} \\
Segment-by-paragraph & 0.9020 & 0.9734 & 0.7818 & 0.8975 & 0.7425 & 0.5837 & \textbf{0.7399} & 0.6483 & -- & -- \\
Equal-length split & 0.8570 & 0.9669 & 0.7587 & 0.8946 & 0.6565 & 0.5018 & 0.6583 & 0.5424 & 0.8585 & 0.2975 \\
TextTiling & 0.8054 & 0.9424 & 0.7376 & 0.8801 & 0.6560 & 0.4842 & 0.6395 & 0.5604 & 0.8642 & 0.2956 \\
ClusterSemanticChunker & 0.8462 & 0.9494 & 0.7499 & 0.8871 & 0.6511 & 0.4540 & 0.6527 & 0.5357 & 0.8607 & 0.3039 \\
LumberChunker (LLM) & 0.8789 & 0.9694 & 0.7700 & 0.9024 & 0.6915 & 0.5064 & 0.6659 & 0.5406 & 0.8601 & 0.3001 \\
\bottomrule
\end{tabular*}
\end{table}

\begin{table}[H]
\caption{\textbf{Overlap setting:} complete-evidence coverage at the two display tiers, same corpora and convention as Table~\ref{tab:main_nonoverlap}.}
\label{tab:main_overlap}
\centering
\scriptsize
\setlength{\tabcolsep}{3pt}
\begin{tabular*}{\textwidth}{@{\extracolsep{\fill}}l cc cc cc cc cc}
\toprule
& \multicolumn{2}{c}{\textbf{Congress report}} & \multicolumn{2}{c}{\textbf{White paper}} & \multicolumn{2}{c}{\textbf{QASPER}} & \multicolumn{2}{c}{\textbf{PeerQA}} & \multicolumn{2}{c}{\textbf{PrivacyQA}} \\
\cmidrule(lr){2-3}\cmidrule(lr){4-5}\cmidrule(lr){6-7}\cmidrule(lr){8-9}\cmidrule(lr){10-11}
\textbf{Method} & @100 & @10\% & @100 & @10\% & @100 & @10\% & @100 & @10\% & @100 & @10\% \\
\midrule
\textbf{\method{} (greedy)} & \textbf{0.9165} & \textbf{0.9781} & \textbf{0.8656} & \textbf{0.9545} & \textbf{0.7412} & \textbf{0.5786} & \textbf{0.7399} & \textbf{0.6483} & \textbf{0.8747} & \textbf{0.3302} \\
Fixed window $L{=}bl$, ov.\ 20\% & 0.8708 & 0.9626 & 0.7266 & 0.8846 & 0.6807 & 0.5296 & 0.6939 & 0.5456 & 0.8581 & 0.3138 \\
Fixed window $L{=}2bl$, ov.\ 20\% & 0.8778 & 0.9609 & 0.7889 & 0.9148 & 0.6554 & 0.4721 & 0.6901 & 0.5672 & 0.8686 & 0.2947 \\
Sentence-Window $\pm1$ & 0.8312 & 0.9538 & 0.7469 & 0.8937 & 0.7086 & 0.5206 & 0.6862 & 0.5555 & 0.8515 & 0.2739 \\
Sentence-Window $\pm2$ & 0.8464 & 0.9672 & 0.7733 & 0.9079 & 0.6830 & 0.5033 & 0.6600 & 0.5436 & 0.8535 & 0.2639 \\
Parent-Document & 0.8615 & 0.9579 & 0.7713 & 0.8909 & 0.7095 & 0.5238 & 0.6830 & 0.5595 & 0.8461 & 0.2945 \\
Auto-Merging (4 sent.) & 0.8232 & 0.9561 & 0.7559 & 0.8947 & 0.6801 & 0.4881 & 0.6590 & 0.5538 & 0.8509 & 0.2650 \\
\bottomrule
\end{tabular*}
\end{table}

\textbf{Non-overlap.} The exact DP leads in $9/10$ displayed cells. Its only loss is PeerQA at $@100$, where the paragraph baseline is ahead by just $0.58$ points and the evidence is often already paragraph-aligned. The largest gains occur on the white paper ($+3.98$ points at $@100$, $+2.61$ at $@10\%$); near-ties on the Congress report and PrivacyQA indicate less room when simple segmentation already matches the evidence geometry.

\textbf{Overlap.} Greedy leads in all $10/10$ cells, by $0.61$--$8.11$ points over the strongest baseline. The largest gaps occur on the white paper at $@100$ and PeerQA at $@10\%$; gains remain positive on PrivacyQA. Since all rows share the same overlap and returned-context conventions, the pattern reflects evidence-aware selection rather than unequal evaluation budgets.

\FloatBarrier
\subsection{Budget Analysis}
\label{sec:budget}

Figure~\ref{fig:covid_budget_curves} traces the white-paper budget range; the other four datasets appear in Appendix~\ref{app:budget_curves}. Coverage rises monotonically for every method, but QESChunker separates at the smallest budgets and remains ahead in both settings. The gap is widest before the curves flatten, when only a small document fraction can be returned.

\begin{figure}[H]
\centering
\includegraphics[width=0.70\linewidth]{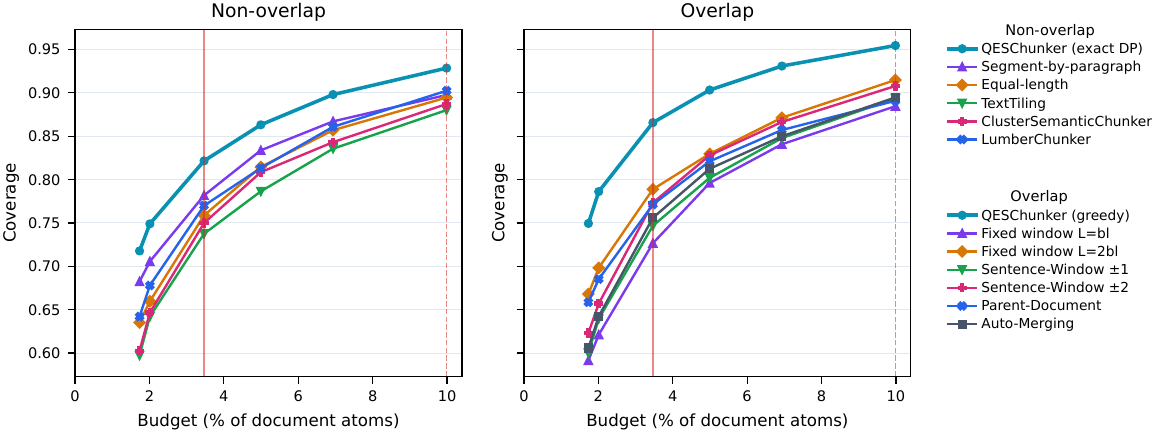}
\caption{Coverage versus returned-atom budget on the white paper. Vertical lines mark $@100$ and $@10\%$.}
\label{fig:covid_budget_curves}
\end{figure}

The appendix curves repeat this pattern with different strength: Congress methods converge near saturation; QASPER and PeerQA retain wider overlap-side gaps at low and medium budgets; and PrivacyQA has tightly clustered non-overlap curves but clear overlap separation. Thus more context helps every method while reducing the marginal value of optimized boundaries near the coverage ceiling.

\FloatBarrier

\section{Limitations}
\label{sec:limitations}

On one hand, our gains are measured at the retrieval level as complete-evidence recall. Providing more complete and effective evidence can plausibly improve downstream generation, but the rate at which recall gains translate into final answer quality remains unknown. This conversion depends on the generator, prompting, context ordering, evidence redundancy, and answer-level evaluation criteria; we therefore do not claim a fixed or proportional end-to-end improvement.

On the other hand, boundary fitting depends heavily on the quality and coverage of existing question--evidence supervision. It is thus best suited to documents that are queried repeatedly and have accumulated sufficient supervision---for example, policy reports, FAQs, or frequently discussed papers. For one-off, cold-start, or sparsely annotated documents, such supervision may be unreliable or unavailable, and performance may be driven mainly by the structural prior and the retriever rather than by learned boundaries.

\section{Conclusion}
\label{sec:conclusion}

QESChunker casts chunk construction as supervised, budget-aware optimization rather than heuristic preprocessing. A shared hit-gated coverage objective yields two complementary solvers: exact dynamic programming for non-overlapping partitions and benefit-cost submodular greedy for overlapping indices. Across five corpora, the two variants improve complete-evidence recall in nearly all displayed settings, with the largest gains appearing at tight returned-context budgets and smaller margins near saturation. The parameter sweeps further show that performance remains stable across broad structural-weight and candidate-length ranges. Together, these results indicate that question--evidence supervision can translate recurring retrieval demand into principled, document-specific static indices under either overlap regime.


\clearpage
\subsection*{AI use statement}
Generative AI was used to generate fitting questions and evidence IDs and to judge auxiliary outputs. Data passed legality, membership, length, hallucination, and duplication checks; public evaluation used only human or expert labels. Judging was anonymized, order-swapped, and conducted per fact. AI was not used for theoretical development or solver implementation. The authors reviewed all outputs and accept responsibility.

\subsection*{Ethics statement}
This work studies document segmentation for retrieval-augmented generation and uses publicly available scientific QA data (QASPER, PeerQA), a public privacy-policy QA corpus (PrivacyQA), and public government policy documents. It involves no human subjects, no private data, and no personal identifiers. The main dual-use consideration is that better evidence retrieval can improve both beneficial and harmful downstream generation; we mitigate fabrication risk in our own evaluation via hallucination screening, and we report negative and boundary results (scope, oracle ceiling, cost accounting) to avoid overstating capability. LLM-annotated evidence on the self-built corpora may inherit annotator-model biases; we flag this as a limitation and rely on human-annotated QASPER and author-annotated PeerQA for our highest-confidence claims.

\subsection*{Reproducibility statement}
The method is fully specified in \S\ref{sec:objective}--\S\ref{sec:greedy}: the objective \eqref{eq:objective} with weights \eqref{eq:weights} and tier ratios \eqref{eq:tiers}, the DP recurrence and predecessor set \eqref{eq:dp}--\eqref{eq:predecessors} with the whole-paragraph exception (Algorithm~\ref{alg:dp}), the NP-hardness reduction (Theorem~\ref{thm:nphard}), and the greedy with its guarantee, lazy evaluation, singleton safeguard (Algorithm~\ref{alg:greedy}, Theorem~\ref{thm:greedy}). The residual-budget concentration term \eqref{eq:concentration}, evaluation convention, and sensitivity analyses are given in \S\ref{sec:main}, \S\ref{sec:setup}, and \S\ref{sec:sweeps}, respectively. Correctness claims are backed by brute-force enumeration for the DP and by the certificate for the greedy.

\bibliography{iclr2027_conference}
\bibliographystyle{iclr2027_conference}

\appendix
\clearpage
\section{Proof of Proposition~\ref{prop:submodular}}
\label{app:proof}

\begin{proof}
Write $f(S)=\fcov(S)+f_{\mathrm{pri}}(S)$ with
$\fcov(S)=\sum_{(j,r)\in\bigcup_{C\in S}\Gamma(C)}w(j,r)$ and $f_{\mathrm{pri}}(S)=\sum_{C\in S}\lambda(b_C)$.

\emph{$\fcov$ is monotone submodular.} Each element $e=(j,r)$ with weight $w(e)\ge0$ is covered by $S$ iff some $C\in S$ has $e\in\Gamma(C)$; thus $\fcov(S)=\sum_e w(e)\,\ind{e\text{ covered by }S}$ is a weighted coverage function, which is monotone (adding chunks cannot uncover) and submodular (the marginal contribution of a chunk is the total weight of elements it covers for the first time, which is non-increasing as $S$ grows).

\emph{$f_{\mathrm{pri}}$ is modular and non-negative.} Each summand $\lambda(b_C)\ge0$ depends only on the single chunk $C$; adding a chunk $C$ to any $S$ increases $f_{\mathrm{pri}}$ by exactly $\lambda(b_C)$, independent of $S$. A modular function is submodular with equality, and monotone since the increment is non-negative.

\emph{Sum.} The sum of two monotone submodular functions is monotone submodular, so $f$ is monotone submodular. The concentration term $f_2$ in \eqref{eq:concentration} is a sum over $j$ of a $\max$ over selected chunks of modular-in-$C$ terms gated by $C\cap\Hj\neq\varnothing$; a non-negative weighted $\max$ over selected sets is monotone submodular, so $f+\delta f_2$ with $\delta\ge0$ preserves monotone submodularity.
\end{proof}

\textbf{On the partition decomposition.} If $S$ is a partition, its chunks are pairwise disjoint, so no element $(j,r)$ can be covered by two different chunks (each block $r$ lies in exactly one chunk). Hence $\bigcup_{C\in S}\Gamma(C)$ is a disjoint union and $\fcov(S)=\sum_{C\in S}g_{\mathrm{cov}}(C)$ with $g_{\mathrm{cov}}(C)=\sum_{(j,r)\in\Gamma(C)}w(j,r)$. Together with the already-additive prior, $f(S)=\sum_{C\in S}g(C)$ is additive over chunks, which is precisely the optimal-substructure exploited by \eqref{eq:dp}.

\clearpage
\section{Full proof of Theorem~\ref{thm:nphard}}
\label{app:nphard}

We use the construction and notation from the proof sketch.

\begin{proof}
\textbf{Forward implication.}
Suppose the grid instance has a cover with at most $k$ horizontal
length-$2$ segments. After encoding the rows on the line, discard any
interval containing no encoded endpoint. Shift every remaining interval
rightward until its left endpoint is its smallest contained encoded point.
This preserves all covered endpoints and gives an integer left endpoint.
We therefore obtain at most $k$ intervals $[t,t+2]$ that contain at least
one point from every pair $\{p_j,p_j+D\}$.

Each normalized $t$ is an encoded endpoint, so
\[
  3\le t\le p_{\max}+D=n-L+1.
\]
Replace each $[t,t+2]$ by $C_t=[t,t+D+2]$. These are legal chunks, their
total cost is at most $kL=B_{\mathrm{index}}$, and the coverage
correspondence gives $\fcov(S)=1$.

\textbf{Reverse implication.}
Suppose $S\subseteq\mathcal V_L$ satisfies the budget and $\fcov(S)=1$.
Remove chunks with $\Gamma(C)=\varnothing$, preserving coverage. Every
remaining chunk covers some question $j$, so it contains $p_j+D$ and at
least one point in $\{p_j,p_j+2D\}$. These positions differ by $D$; hence
each such chunk costs at least $D+1$. Since $k\le m$ and $D=q+2m+3$,
\[
  (k+1)(D+1)-kL=D+1-2k>0.
\]
Thus at most $k$ nonempty-coverage chunks remain.

For each remaining chunk $[a,b]$, form $C_a=[a,a+D+2]$. This extension stays
inside the document because the original chunk contains an evidence block,
so
\[
  a\le p_j+D\le p_{\max}+D=n-L+1.
\]
Since $b-a+1\le L$, $C_a$ contains the original chunk and preserves all its
covered pairs. The resulting family has at most $k$ length-$L$ chunks and
still covers every question. The coverage correspondence now shows that
the intervals $[a,a+2]$ contain at least one point from every pair
$\{p_j,p_j+D\}$. As encoded points on distinct rows are more than $2$
apart, these intervals map back to at most $k$ horizontal length-$2$
segments covering the grid instance.

\textbf{Polynomial size and NP membership.}
Because $q=O(m)$, we have $D,L=O(m)$ and $n=O(m^2)$. The full candidate
family has size $O(nL)=O(m^3)$, so the reduction is polynomial. A selected
family is a polynomial-size certificate, and its cost and coverage are
polynomially checkable. This proves NP-completeness.

If candidates are restricted to length exactly $L$, the forward direction
is unchanged. In the reverse direction, the budget immediately permits at
most $k$ chunks, so the short-chunk argument is unnecessary.
\end{proof}

\clearpage
\section{Additional Budget Curves}
\label{app:budget_curves}

We report the full returned-context budget curves for the four datasets not shown in Figure~\ref{fig:covid_budget_curves}. Each panel follows the same non-overlap/overlap convention as the white-paper analysis.

\begin{figure}[H]
\centering
\includegraphics[width=0.92\linewidth]{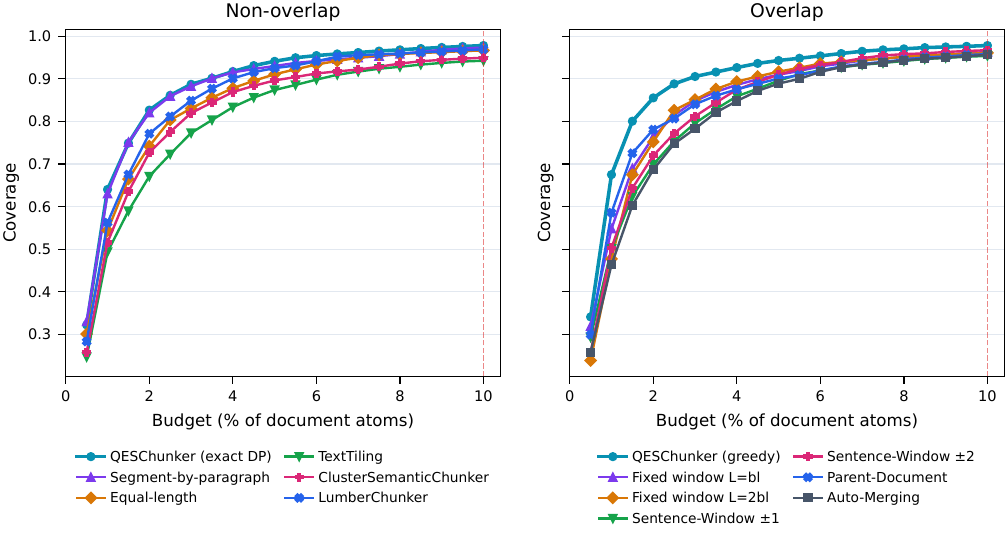}
\caption{Budget curves on the Party Congress report. Left: non-overlap; right: overlap.}
\label{fig:congress_budget_curves}
\end{figure}

\begin{figure}[H]
\centering
\includegraphics[width=0.92\linewidth]{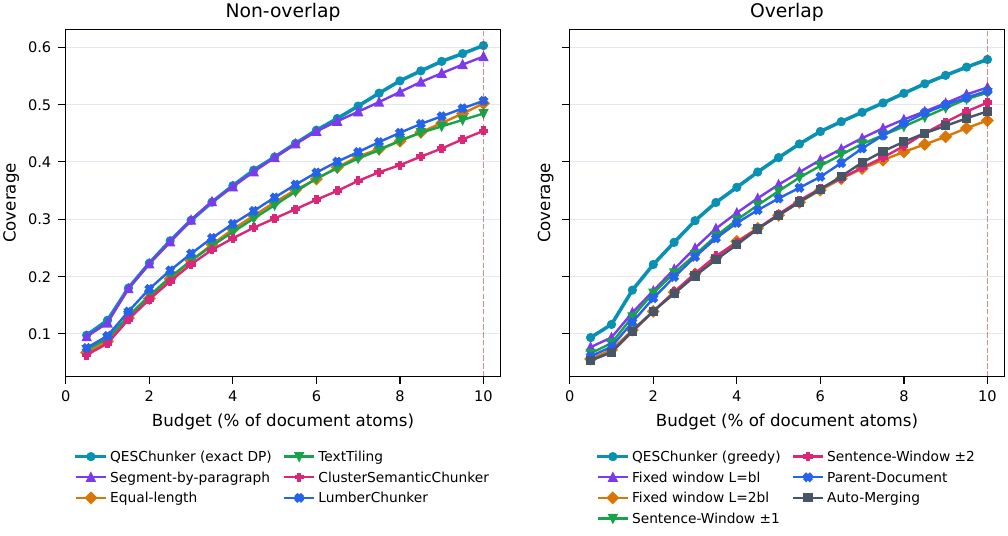}
\caption{Budget curves on QASPER. Left: non-overlap; right: overlap.}
\label{fig:qasper_budget_curves}
\end{figure}

\clearpage
\begin{figure}[H]
\centering
\includegraphics[width=0.92\linewidth]{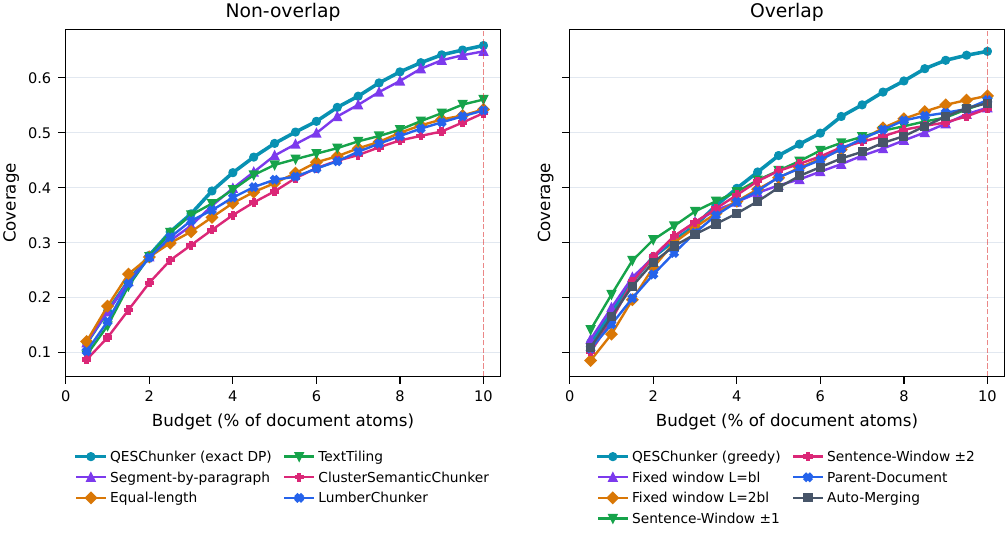}
\caption{Budget curves on PeerQA. Left: non-overlap; right: overlap.}
\label{fig:peerqa_budget_curves}
\end{figure}

\begin{figure}[H]
\centering
\includegraphics[width=0.92\linewidth]{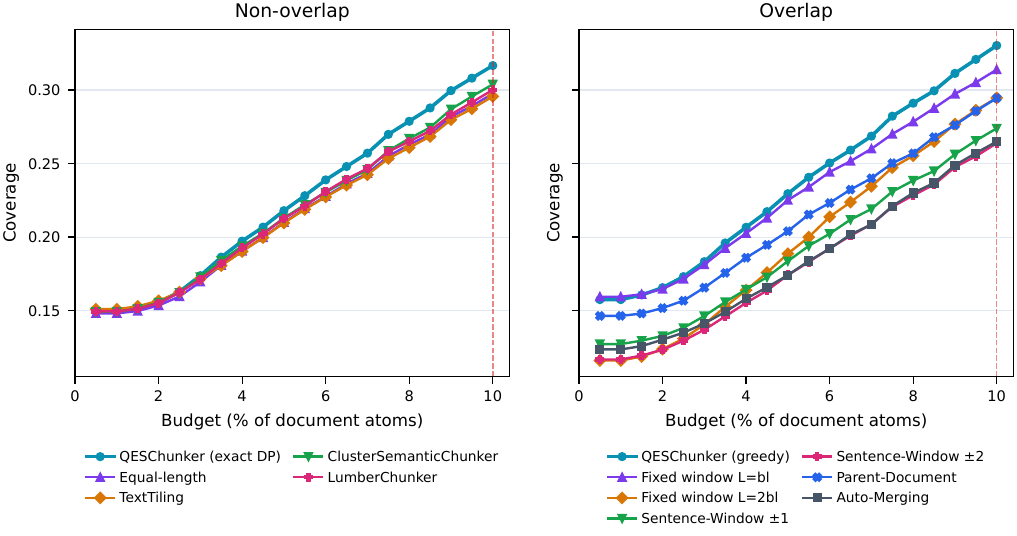}
\caption{Budget curves on PrivacyQA. Left: non-overlap; right: overlap.}
\label{fig:privacyqa_budget_curves}
\end{figure}

\end{document}